\documentclass[sigconf]{acmart}

\AtBeginDocument{%
  \providecommand\BibTeX{{%
    \normalfont B\kern-0.5em{\scshape i\kern-0.25em b}\kern-0.8em\TeX}}}

\copyrightyear{2023}
\acmYear{2023}
\setcopyright{rightsretained}
\acmConference[SIGCSE 2023]{Proceedings of the 54th ACM Technical Symposium on Computing Science Education V. 2}{March 15--18, 2023}{Toronto, ON, Canada}
\acmBooktitle{Proceedings of the 54th ACM Technical Symposium on Computing Science Education V. 2 (SIGCSE 2023), March 15--18, 2023, Toronto, ON, Canada} 
\acmDOI{10.1145/3545947.3569630} 
\acmISBN{978-1-4503-9433-8/23/03}

\usepackage{todonotes}

\begin{document}

\title{Automatically Generating CS Learning Materials with Large Language Models}


\author{Stephen MacNeil}
\affiliation{%
  \institution{Temple University}
  \streetaddress{1801 N Broad St}
  \city{Philadelphia}
  \state{PA}
  \country{USA}
  \postcode{19122}
}
\email{stephen.macneil@temple.edu}
\orcid{0000-0003-2781-6619}

\author{Andrew Tran}
\affiliation{%
  \institution{Temple University}
  \streetaddress{1801 N Broad St}
  \city{Philadelphia}
  \state{PA}
  \country{USA}
  \postcode{19122}
}
\email{andrew.tran10@temple.edu}

\author{Juho Leinonen}
\affiliation{%
  \institution{Aalto University}
  \city{Espoo}
  \country{Finland}}
\email{juho.2.leinonen@aalto.fi}
\orcid{0000-0001-6829-9449}

\author{Paul Denny}
\affiliation{%
  \institution{The University of Auckland}
  \city{Auckland}
  \country{New Zealand}}
\email{paul@cs.auckland.ac.nz}
\orcid{0000-0002-5150-9806}

\author{Joanne Kim}
\affiliation{%
  \institution{Temple University}
  \streetaddress{1801 N Broad St}
  \city{Philadelphia}
  \state{PA}
  \country{USA}
  \postcode{19122}
}
\email{joanne.kim@temple.edu}

\author{Arto Hellas}
\affiliation{%
  \institution{Aalto University}
  \city{Espoo}
  \country{Finland}}
\email{arto.hellas@aalto.fi}
\orcid{0000-0001-6502-209X}

\author{Seth Bernstein}
\affiliation{%
  \institution{Temple University}
  \streetaddress{1801 N Broad St}
  \city{Philadelphia}
  \state{PA}
  \country{USA}
  \postcode{19122}
}
\email{seth.bernstein@temple.edu}

\author{Sami Sarsa}
\affiliation{%
  \institution{Aalto University}
  \city{Espoo}
  \country{Finland}}
\email{sami.sarsa@aalto.fi}
\orcid{0000-0002-7277-9282}

\renewcommand{\shortauthors}{MacNeil, et al.}

\begin{abstract}

Recent breakthroughs in Large Language Models (LLMs), such as GPT-3 and Codex, now enable software developers to generate code based on a natural language prompt. Within computer science education,  researchers are exploring the potential for LLMs to generate code explanations and programming assignments using carefully crafted prompts. These advances may enable students to interact with code in new ways while helping instructors scale their learning materials. 
However, LLMs also introduce new implications for academic integrity, curriculum design, and software engineering careers. This workshop will demonstrate the capabilities of LLMs to help attendees evaluate whether and how LLMs might be integrated into their pedagogy and research. We will also engage attendees in brainstorming to consider how LLMs will impact our field.
 
\end{abstract}




\maketitle

\section{Introduction}

Educational technology can have a transformational effect on teaching and learning in computer science classrooms. Intelligent tutoring systems provide students with real-time formative feedback on their work to help them get unstuck~\cite{10.1145/3304221.3319759} when peers and instructors are not available. Online tutorials and videos have enabled instructors to `flip their classes' and consider new methods for content delivery~\cite{giannakos2014reviewing, 10.1145/2787622.2787709}, making necessary space for active learning and collaboration during class time. Anchored collaboration~\cite{kolodner1998integrating, dorn2015piloting} and subgoal labeling~\cite{kim2013learnersourcing, decker2020using} have created more engaging online learning spaces where students co-construct their knowledge and build on each other's ideas. Clicker quizzes and peer instruction methods enable instructors to evaluate students' misconceptions within large classes in real-time~\cite{crouch2001peer}. Technology advances and education technology can not only improve classroom experiences but also create new models and opportunities for teaching and learning.

Large language Models (LLMs) are similarly poised to impact computer science classrooms. LLMs are machine learning models that are trained on a large amount of text data. These models are designed to learn the statistical properties of language in order to predict the next word in a sequence or generate new text. LLMs are capable of natural language understanding and text generation which enables many use cases ranging from creative story writing~\cite{yuan2022wordcraft} to using LLMs to write about themselves~\cite{thunstrom2022can}. In computer science classroom settings, LLMs have the potential to provide high-quality code explanations for students at scale~\cite{macneil2022generating, macneil2023experiences, sarsa2022automatic}.

In this workshop, we will demonstrate LLMs' capabilities to inspire instructors and researchers to consider how this new technology might integrate with their existing pedagogy. 
We also will discuss the potential impacts that LLMs might have on curricula and students' careers. Given that LLMs can generate code based on a natural language prompt, the skills and job requirements may change for software engineers. Software engineers may take on more design-oriented roles and serve as software architects while LLMs write (most of) the source code. This might lead to courses that focus on prompt engineering, code evaluation, and debugging.

\begin{figure}
    \centering
 \includegraphics[width=0.85\linewidth]{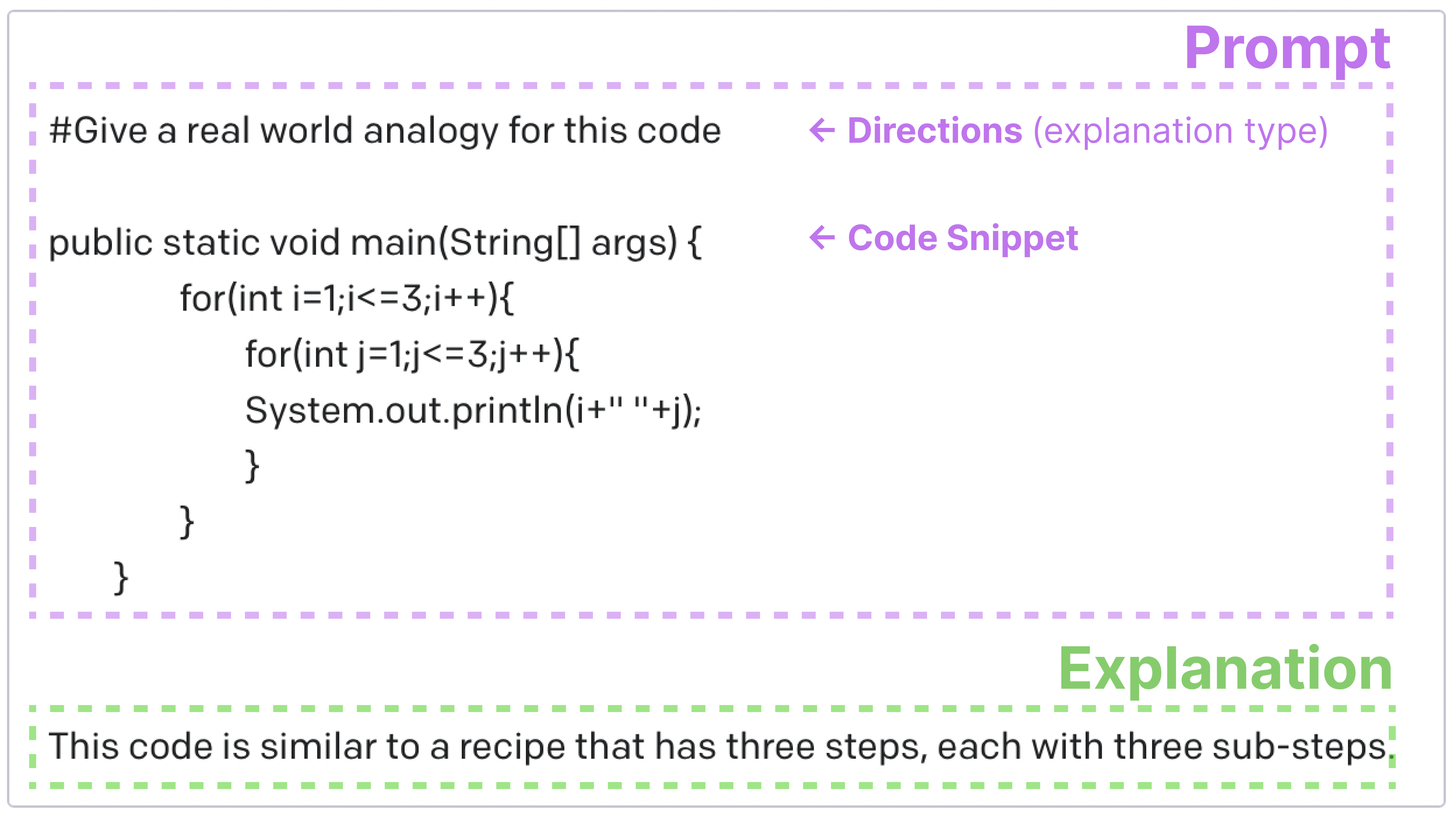}
    \caption{An example code snippet with an explanation generated using GPT-3.}
\label{fig:my_label}
\end{figure}

\begin{figure*}
    \centering
    \includegraphics[width=0.75\linewidth]{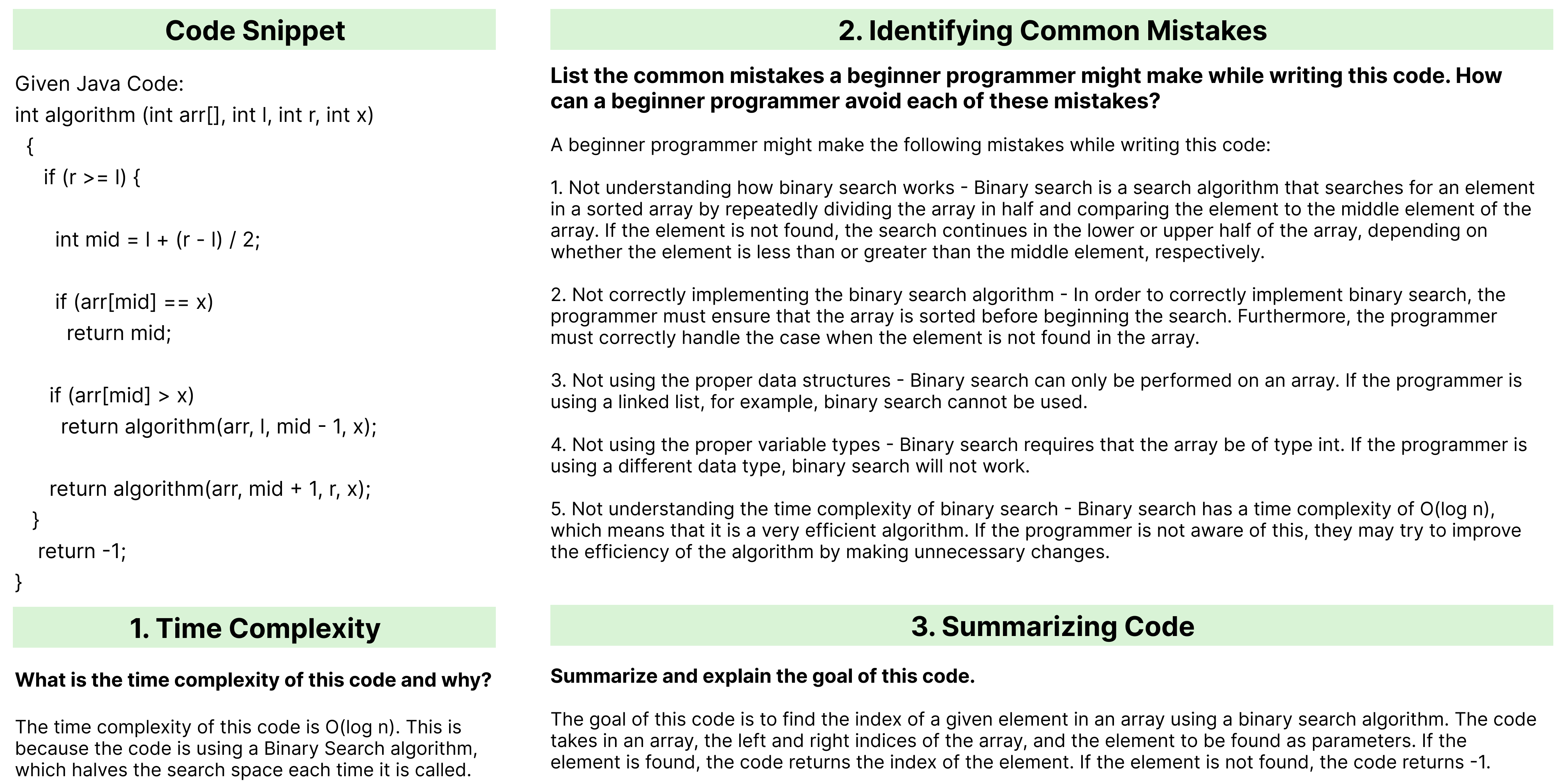}
    \caption{An example code snippet with an analogy-based explanation generated.}
\label{fig:my_label}
\end{figure*}

\subsection{Generating Explanations of Code}

High-quality code explanations enable students to better understand code and efficiently learn programming concepts~\cite{Marwan2020Adaptive}. Common automated methods to explain code snippets and coding concepts include tracing program execution ~\cite{guo2013online}, defining terms~\cite{head2015tutorons}, providing hints~\cite{price2017isnap}, and presenting feedback on errors~\cite{Marwan2020Adaptive, price2017isnap}.  
These techniques either leverage heuristics which limits their generalizability or rely on instructors to manually configure and pregenerate content. However, LLMs have the potential to scale these efforts and generalize to multiple languages and programming contexts. Our research team has developed code-explanations with the LLMs Codex~\cite{sarsa2022automatic} and GPT-3~\cite{macneil2022generating, macneil2023experiences} which resulted in the development of a design space for LLM-generated code explanations~\cite{macneil2022generating}.

\subsection{Generating Assignments}

Students benefit from frequent hands-on 
practice with programming assignments~\cite{7126236}. Assignments are most engaging when they are personalized toward a student's personal interests~\cite{haungs2012improving} and when they provide sufficient instructions and examples. However, it is time-consuming to create and maintain high-quality assignments. Previous researchers have techniques to automatically generate assignments, but they require instructors to build and maintain  templates~\cite{7166637}. To provide high-quality assignments at scale, our team has developed prompts to generate programming assignments using OpenAI Codex~\cite{sarsa2022automatic}. 
Based on these prior experiences, we will share best practices with attendees. 

\subsection{Generating Code}

Large language models have the potential to change the roles and responsibilities of software developers. For example, Git\-Hub's Copilot can generate code for programmers based on natural language prompts~\cite{chen2021evaluating}. The generated code is high enough quality to lead researchers to raise concerns about cheating~\cite{Finnie-Ansley2022Robots}. This ability to generate high-quality code may affect software engineering jobs. Engineers may be expected to elicit code requirements, write prompts, and debug the resulting code. In the workshop, we will explore how LLMs may affect the way we prepare our students.  

\section{Workshop Attendees}

Our workshop is designed primarily with educators and researchers in mind; however, we plan to encourage student attendees at SIGCSE to participate in our workshop and share their perspectives. Our research team consists of faculty, researchers, and undergraduate students to provide a balanced perspective and to make the workshop welcoming to attendees at various points in their careers. We have considered additional methods to make our workshop an inclusive space. We greet attendees and  create space for them to share their names and the pronouns they use. We will also provide building information including the nearest gender-neutral bathroom, elevator, and quiet rooms to reduce barriers to participation.

\section{Schedule}

The goal of our workshop is to give participants an awareness of the capabilities of LLMs to support their pedagogy, to get practice using LLMs and learn best practices in prompt engineering, and to brainstorm with their colleagues the ways large language models can support their pedagogy. 

\begin{itemize}

    \item \textbf{Pre-workshop Activities:} We will share a tutorial to guide attendees through creating Github Copilot and OpenAI accounts with sample prompts to try on their own before the workshop. Free credits are currently available. 
    \item \textbf{Introductions (20 mins):} The team and attendees introduce themselves. Attendees will engage in a speed dating activity to get to know others one-on-one.
    \item \textbf{Demonstration (10 mins):} Our team will demonstrate the capabilities of GPT-3 (supported by materials). 
    \item \textbf{Guided Activity 1 (20 mins):} Participants will work in pairs to solve programming assignments with Github Copilot. 
    \item \textbf{Break (10 mins)}
    \item \textbf{Guided Activity 2 (25 mins):} Participants will work in pairs to generate code explanations using GPT-3. 
    \item \textbf{Guided Activity 3 (25 mins):} Participants will work in pairs to create programming assignments using Codex.
    \item \textbf{Break (10 mins)}
    \item \textbf{Group brainstorming (25 mins):} As a think-pair-share activity, attendees will work in pairs on a shared Miro board to brainstorm ideas for integrating LLMs into their courses.
    \item \textbf{Exploratory learning (25):} attendees will use our resources to further explore LLMs, explore ways to realize brainstormed ideas, and explore LLMs beyond the workshop activities, including testing unique prompt ideas. 
    \item \textbf{Debrief (10 mins):} Summarization of the workshop and key insights, initiating collaborations, etc.
\end{itemize}

\section{Disseminating Workshop Results}

Our workshop team will create a website leading up to the workshop which will host and maintain resources on using LLMs in CS classrooms. After the workshop, we will update the website with content from the Miro boards and a joint reflection written by workshop organizers on the challenges and opportunities identified by attendees. This idea is inspired by existing websites with advice for CS Teaching (e.g.: \textit{https://www.csteachingtips.org/}).

\section{Organizers}

Our team has extensive experience using LLMs to write code~\cite{Finnie-Ansley2022Robots}, generate code explanations~\cite{macneil2023experiences, macneil2022generating}, and create assignments~\cite{ sarsa2022automatic}. Our team includes three faculty members, two researchers, and three undergraduate students who have all published research on using LLMs for CS Education. We strongly believe that this diverse convergence of faculty, researchers, and students will be essential to ensure that LLMs have the most positive potential impact on computing education.

\balance
\bibliographystyle{ACM-Reference-Format}
\bibliography{sample-base}

\end{document}